\newcounter{myctr}

\documentclass{ws-ijqi}

\begin{document}

\markboth{A. Allevi, S. Olivares, M. Bondani} {Experimental quantification of non-Gaussianity...}


\title{EXPERIMENTAL QUANTIFICATION OF NON-GAUSSIANITY OF PHASE-RANDOMIZED COHERENT STATES}

\author{ALESSIA ALLEVI}

\address{Dipartimento di Scienza eAlta Tecnologia, Universit\`a degli Studi dell'Insubria, and\\ C.N.I.S.M., U.d.R. Como, via Valleggio 11, Como, 22100, Italy.\\
alessia.allevi@uninsubria.it}

\author{STEFANO OLIVARES}

\address{Dipartimento di Fisica, Universit\`a degli Studi
di Milano, and\\ C.N.I.S.M., U.d.R. Milano Statale, via Celoria 16, 20133 Milano, Italy.\\
stefano.olivares@fisica.unimi.it}

\author{MARIA BONDANI}

\address{Istituto di Fotonica e Nanotecnologie, CNR, and\\ C.N.I.S.M., U.d.R. Como, via Valleggio 11, Como, 22100, Italy.\\
maria.bondani@uninsubria.it}

\maketitle


\begin{abstract}
We present the experimental investigation of the non-Gaussian nature of some mixtures of Fock states by reconstructing their Wigner function and exploiting two recently introduced measures of non-Gaussianity. In particular, we demonstrate the consistency between the different approaches and the monotonicity of the two measures for states belonging to the class of phase randomized coherent states. Moreover, we prove that the exact behavior of one measure with respect to the other depends on the states under investigation and devise possible criteria to discriminate which measure is more useful for the characterization of the states in realistic applications.
\end{abstract}


\section{Introduction}    

Gaussian states, \emph{i.e.} states with Gaussian Wigner functions, are the key ingredient of many continuous variable (CV) Quantum Information protocols.\cite{RMP:12,oli:rev} However, in order to achieve some relevant tasks, non-Gaussianity (nonG) in the form of non-Gaussian states (states endowed with a non-Gaussian Wigner function) or non-Gaussian operations is either required or desirable. For instance, it has been recently demonstrated that nonG can be used to improve teleportation, cloning and storage; in addition non-Gaussian operations are interesting for the realization of entanglement distillation and noiseless amplification.\cite{barbieri10}
\par
Several implementations of non-Gaussian states have been reported so far, in particular from squeezed light,\cite{lvovsky01,zavatta04} close-to-threshold parametric oscillators\cite{dauria05} and in superconducting circuits.\cite{hofheinz08} Such states have been mainly achieved in the low energy regime\cite{ourjoumtsev07,takahashi10,braczyk10} by using single-photon detectors, visible light photon counters\cite{waks06} and time-multiplexed photo-resolving detectors.\cite{osullivan08} More recently, we extended  the investigation to the mesoscopic regime by exploiting the linear response of hybrid photodetectors\cite{allevi10a,allevi10b,spie12a} and Si-photomultipliers.\cite{spie12b}
\par
Being it recognized as a resource for CV Quantum Information, the need of quantifying the non-Gaussian character of states and operations naturally arises and different non-Gaussianity measures have been proposed.\cite{genoni10} Indeed, not all of them are characterized by an operational meaning. Moreover, in the realistic situations, non-Gaussianity measures based only on quantities that can be experimentally accessed are desirable. To this aim, here we present an experimental work in which we compare the two measures introduced in Refs.~[18] and~[19] by testing them on phase-randomized coherent states or phase-averaged coherent states (PHAVs), a class of states exploited in communication channels and in decoy-state-based quantum key distribution protocols.\cite{OE12} The two measures, being in perfect agreement with each other, can be considered a useful tool to quantify the non-Gaussian nature exhibited by the Wigner functions and testified by the experimental results. Moreover, the sensitivity of such non-Gaussian measures to small changes in the mean number of photons suggests their possible exploitation to test decoy states based on the class of PHAVs.

\section{Quantifying nonG amount}
\label{sec:theory}
A single-mode  PHAV $\varrho_\beta$ is a classical state obtainable by randomizing the phase $\phi$ of a coherent state $| \beta \rangle$, $\beta = |\beta|\,e^{i\phi}$. It is characterized by a density matrix diagonal in the photon-number basis and by a Poissonian photon statistics. The state is obviously phase-insensitive and its Wigner function reads:\cite{bondani09a} 
\begin{equation}
W_{\rm PHAV}(\alpha;\beta) = \int_{0}^{2 \pi} \frac{d \phi}{2 \pi}\, e^{-|\alpha - \beta|^2}= \frac{2}{\pi} \, I_0(4|\alpha||\beta|)\, \exp[-2(|\alpha|^2+|\beta|^2)]\,,
\label{eq:wignerPHAV}
\end{equation}
where $I_0(z)$ is the modified Bessel function. This function, being it endowed with a dip in the origin of the phase space, is clearly non-Gaussian.
\par
Another state exhibiting a nonG nature, which can be useful for application to passive decoy state quantum key distribution,\cite{curty09} can be obtained from the interference of two PHAVs $\varrho_{\beta}$ and $\varrho_{\tilde{\beta}}$ (2-PHAV hereafter) at a beam splitter (BS) with transmissivity $t$. The states outgoing the BS are still diagonal in the photon number basis and the transmitted one is characterized by the following Wigner function:
\begin{equation} \label{eq:wigner2PHAV:B}
W_{\rm 2-PHAV}(\alpha;\beta,\tilde{\beta},t) = \int_{0}^{2 \pi} \frac{d\tilde{\phi}}{2\pi}\,
W_{\rm PHAV}(\alpha-\tilde{\beta}\sqrt{1-t};\beta\sqrt{t})
\end{equation}
where $\tilde{\beta} = |\tilde{\beta}|\,e^{i \tilde{\phi}}$ and the function in the integral is given by Eq.~(\ref{eq:wignerPHAV}). Obviously, the Wigner function of the reflected mode can be obtained by replacing $t$ with the reflectivity $(1-t)$. 
In both cases, the quantification of nonG amount can be achieved by considering two competitive measures recently introduced. 
One of them is based on the Hilbert-Schmidt distance from a Gaussian reference state, namely:
\begin{equation}\label{nonGA}
\delta_{\rm A}[\varrho_{\beta}] = \frac{D^2_{\rm HS}[\varrho_{\beta}, \sigma]}{\mu[\varrho_{\beta}]}=\frac{\mu[\varrho_{\beta}]+\mu[\sigma]+2\kappa[\varrho_{\beta}, \sigma]}{2\mu[\varrho_{\beta}]},
\end{equation}
where $\mu[\varrho]$ is the purity of the state $\varrho$, $\sigma$ is a reference Gaussian state with the same covariance matrix as the state $\varrho_{\beta}$ under investigation and $\kappa[\varrho_{\beta}, \sigma]=$Tr$[\varrho_{\beta} \sigma]$ denotes the overlap between $\varrho_{\beta}$ and $\sigma$.\cite{genoni07}
\par
The second measure we address is given by:
\begin{equation}\label{nonGB}
\delta_{\rm B}[\varrho_{\beta}] = S(\sigma) - S(\varrho_{\beta}),
\end{equation}
where $S(\varrho) = -\hbox{Tr} [\varrho \ln \varrho]$ is the von Neumann entropy of the state $\varrho$.\cite{genoni08} 
As both PHAV and 2-PHAV are diagonal in the photon-number basis, their reference state is a thermal equilibrium state, with the same mean number of photons $N=|\beta|^2$. Moreover, the two measures result based only  on quantities that can be experimentally accessed by direct detection, since both Eqs.~(\ref{nonGA}) and ~(\ref{nonGB}) can be expressed in terms of photon-number distributions.\cite{OE12}
In particular, Eq.~(\ref{nonGA}) reduces to 
\begin{equation}\label{nonGA1}
\delta_{\rm A}[\varrho_\beta] = \frac{1}{2}\left[1-\frac{\sum_n \tau_n (2 p_n- \tau_n)}{\sum_n {p_n}^2}\right],
\end{equation}
where $\tau_n = N^n/(1+N)^{n+1}$ is the photon-number distribution of a single-mode thermal state having $N$ mean number of photons and $p_n$ is the statistics of $\varrho_{\beta}$. In particular, PHAV is described by a Poissonian distribution 
\begin{equation}
p^{\rm PHAV}_n = \exp{(-|\beta|^2)}\, |\beta|^{2n}/n!,\nonumber
\end{equation}
whereas 2-PHAV is characterized by a non-trivial 2-peaks distribution:\cite{zambra07,bondani09b} 
\begin{eqnarray}
 \label{eq:phaseaver}
 p^{\rm 2-PHAV}_{n} &=& \frac{A^{n}}{n!}e^{-A}\sum_{k=0}^n \left(
 \begin{array}{c}
   n  \\
   k  \\
 \end{array} \right)
 \frac{\left(-1\right)^k }{2\pi}\left(\frac{B}{A}\right)^k\frac{\Gamma\left(1/2+ k/2\right)
 \Gamma\left(1/2\right)}{\Gamma\left(1+ k/2\right)} \\ \nonumber
 &\mbox{}&\times {}_1F_2 \left[\left\{1/2+k/2\right\},\left\{1/2,1+k/2\right\},B^2/4 \right]\; ,
\end{eqnarray}
in which $A = |\beta|^2+|\tilde{\beta}|^2$, $B=2 |\beta||\tilde{\beta}|$ and
$_1F_2(a,b,z)$ is the generalized hypergeometric function.\\
On the other hand, Eq.~(\ref{nonGB}) becomes
\begin{equation}\label{nonGB1}
\delta_{\rm B}[\varrho_\beta] = (N+1) \ln (N+1)-N \ln N +\sum_n p_n \ln p_n.
\end{equation}
\section{Experimental results and discussion}
\label{sec:experiment}
We generated the class of PHAVs by exploiting the second harmonics (@ 523 nm, 5-ps pulses) of a mode-locked Nd:YLF laser
amplified at 500 Hz (High-Q Laser Production).
\begin{figure}[htbp]
\centering\includegraphics[width=0.65\textwidth]{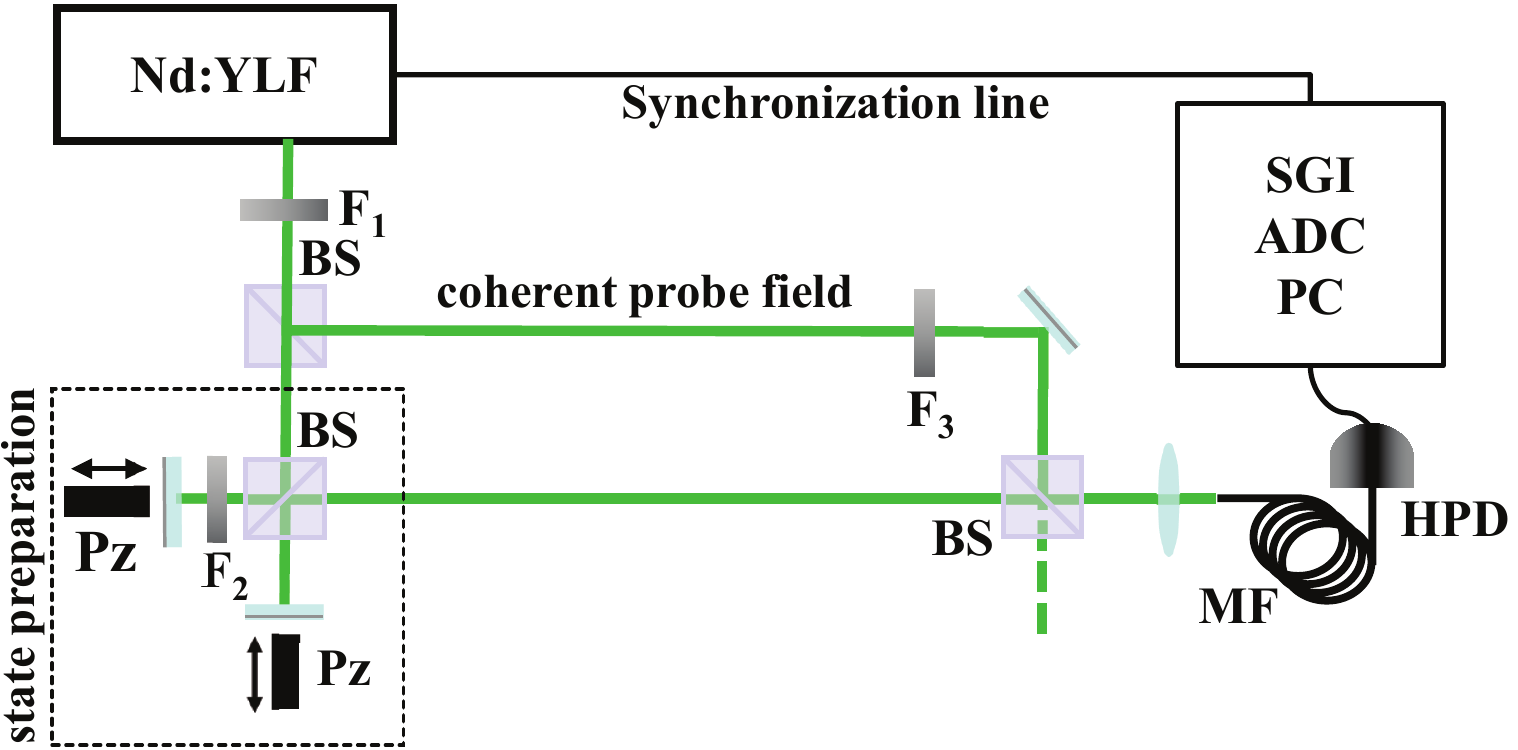}
\caption{Experimental setup. F$_{\rm j}$: variable neutral density filter; BS: 50/50 beam splitter; Pz: piezoelectric movement; MF: multimode fiber (600~$\mu$m core).} \label{setup}
\end{figure}
According to the experimental setup sketched in Fig.~\ref{setup}, we obtained the single PHAV by sending the light pulses to a mirror mounted on a piezo-electric movement. Its displacement, which was controlled by a function generator, was operated at a frequency of 100~Hz and covered a 12~$\mu$m span. Moreover, we produced the 2-PHAV from the interference of two single PHAVs at a BS. A continuous variable density filter F$_1$ allowed us to change the total energy of the states, whereas a second filter F$_2$, inserted in the path of one of the two PHAVs, was used to change the balancing between the two components of the 2-PHAV. 
As the states to be characterized can be fully described by their photon-number distributions, we implemented a direct detection scheme involving a photon-counting detector, $i.e.$ a hybrid photodetector (HPD, R10467U-40, maximum quantum efficiency $\sim$0.5 at 500 nm, Hamamatsu). This detector is characterized not only by a partial photon-counting capability, but also by a linear response up to 100 photons. Thanks to its properties, the HPD can actually operate in the mesoscopic domain, where the states are robust with respect to losses.  The output of the detector was amplified (preamplifier A250 plus amplifier A275, Amptek), synchronously integrated (SGI, SR250, Stanford) and digitized (AT-MIO-16E-1, National Instruments). The gain of the detection apparatus was obtained in a self-consistent way without any $a$ $priori$ calibration.\cite{bondani09c,andreoni09} This method allows us to reconstruct the detected photons distributions of the states, which represent the basic element to retrieve their Wigner function.
Such a goal can be achieved by mixing at a BS the state to be characterized with a coherent probe field whose amplitude and phase are continuously varied.\cite{cahill69,banas96,vogel96} In this case, as both the states are phase-insensitive, we actually reconstruct only a section of their Wigner distribution for fixed phase.
\par
\begin{figure}[htbp]
\centering\includegraphics[width=0.49\textwidth]{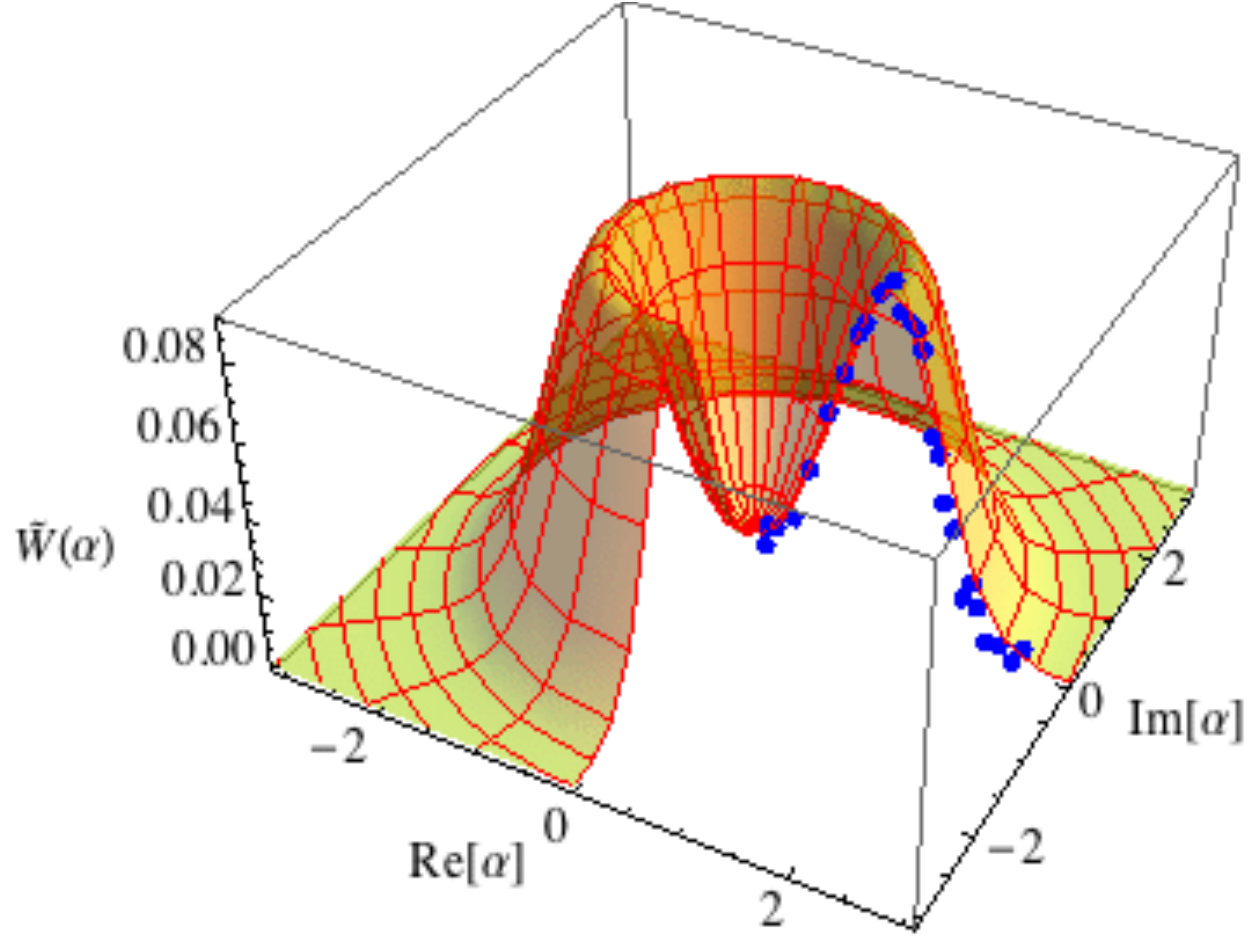} \centering\includegraphics[width=0.49\textwidth]{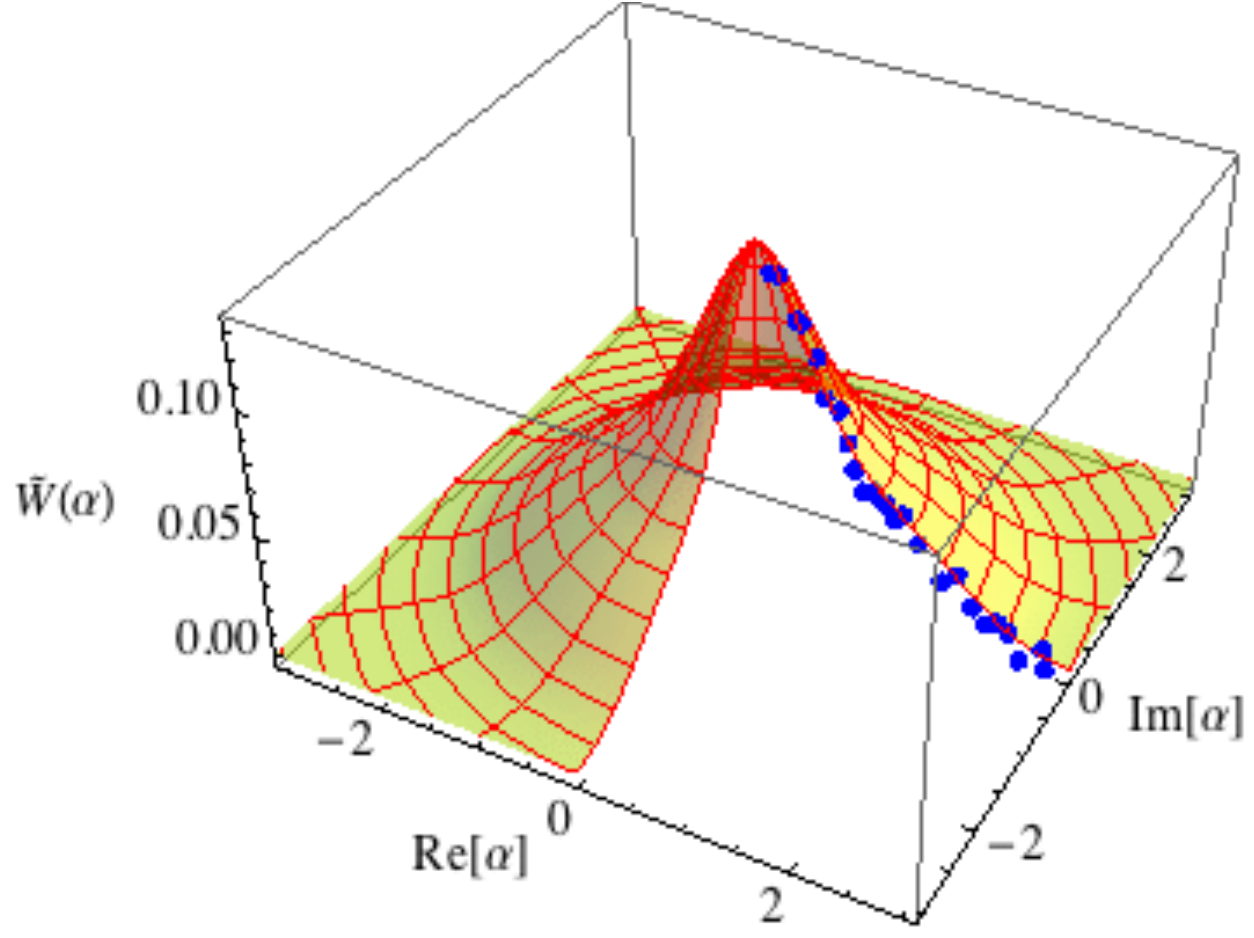}
 \caption{Experimental reconstruction of a section of the phase-insensitive Wigner function of a PHAV (left panel), with $|\beta|^2=1.97$ and $\xi=0.999$ and of a 2-PHAV (right panel), with $|\beta|^2=1.03$, $|\tilde{\beta}|^2=0.91$, $\xi_{\rm P}=0.95$ and $\xi_{\rm S}=1$. Blue dots: experimental data; mesh: theoretical surface.} \label{wig-comp}
\end{figure}
In Fig.~\ref{wig-comp} we plot the experimental data (blue dots) of a single PHAV (left panel) and of a balanced 2-PHAV (right panel), endowed with nearly the same mean number of detected photons, $M_{\rm T}=1.97$ and $M_{\rm T}=1.94$, respectively. In each panel we also show the 3D-theoretical expectations (mesh) for the PHAV and 2-PHAV, respectively:\cite{bondani09a}
\begin{equation}
\tilde{W}_{\rm PHAV}(\sqrt{\xi} \alpha) = W_{\rm PHAV}(\sqrt{\xi} \alpha) e^{-\sqrt{1-\xi}(|\alpha|+|\beta|)}, \label{eq:wignerPHAVoverlap}
\end{equation}
$\xi$ being the overall (spatial and temporal) overlap between the probe and the PHAV, and:
\begin{equation}
\tilde{W}_{\rm 2-PHAV}(\sqrt{\xi_{\rm P}} \alpha) = W_{\rm 2-PHAV}(\sqrt{\xi}_{\rm P} \alpha) e^{-[\sqrt{1-\xi_{\rm P}}|\alpha|+\sqrt{1-\xi_{\rm S}}(|\beta|+|\tilde{\beta}|)]}, \label{eq:wigner2PHAVoverlap}
\end{equation}
where $\xi_{\rm P}$ describes the overall overlap between the probe and the 2-PHAV and $\xi_{\rm S}$ the overall overlap between the two components of the 2-PHAV. In Eqs.~(\ref{eq:wignerPHAVoverlap}) and (\ref{eq:wigner2PHAVoverlap}), $|\beta|^2$ and $|\tilde{\beta}|^2$ are now the mean numbers of photons we measured (see Fig.~\ref{wig-comp}), thus including the quantum efficiency. In fact, it is worth noting that for classical states the functional form of the Wigner function is preserved also in the presence of losses and its expression, given in terms of detected photons, reads $ \tilde{W}(\alpha) =\frac{2}{\pi} \sum_{m=0}^\infty (-1)^m p_{m,\alpha}^{\ el}$, where $p_{m,\alpha}^{\ el}$ represent the detected-photon distributions of the state to be measured displaced by the probe field.\cite{bondani09a} As testified by the very high values of the overlaps $\xi$, $\xi_{\rm P}$ and $\xi_{\rm S}$ reported in the caption of Fig.~\ref{wig-comp}, we actually achieved a very good superposition in aligning the system.
From a direct comparison between the two panels it emerges that the states under investigation are non-Gaussian. In fact, the Wigner function of a single PHAV is characterized by a dip in the origin of the phase space, whereas that of a 2-PHAV with almost the same mean value exhibits a peak in the origin followed by a ``shoulder''.
Moreover, it is worth noting that the measurements were actually performed in the mesoscopic photon-number domain, as the reconstruction of the Wigner functions was achieved by displacing either the PHAV or the 2-PHAV with a coherent field whose intensity was changed from zero up to four times the mean value of the states themselves.
\par
To quantify the nonG amount, we considered the measures introduced in Sec.~\ref{sec:theory}. As from the experimental point of view we do not have access to photons, we calculated similar expressions, $\epsilon_{\rm A}$ and $\epsilon_{\rm B}$, for detected photons, which represent lower bounds to nonG.\cite{genoni10,allevi10a} In particular we found $\epsilon_{\rm A} = 0.207 \pm 0.004$ and $\epsilon_{\rm B} = 0.156 \pm 0.020$ for the single PHAV, whereas we obtained $\epsilon_{\rm A} = 0.036 \pm 0.005$ and $\epsilon_{\rm B} = 0.012 \pm 0.025$ for the 2-PHAV. The consistency between the two measures, together with the fact that measure $\epsilon_{\rm A}$ can be directly expressed in terms of Wigner functions\cite{genoni07}, demonstrate that a Wigner function exhibiting a dip in the origin of the phase
space is more non-Gaussian than one characterized by a peak in the origin followed by a ``shoulder''. Moreover, the results prove that combining two non-Gaussian states does not necessarily lead to an increase of the overall nonG. 
\begin{figure}[htbp]
\centering\includegraphics[width=0.49\textwidth]{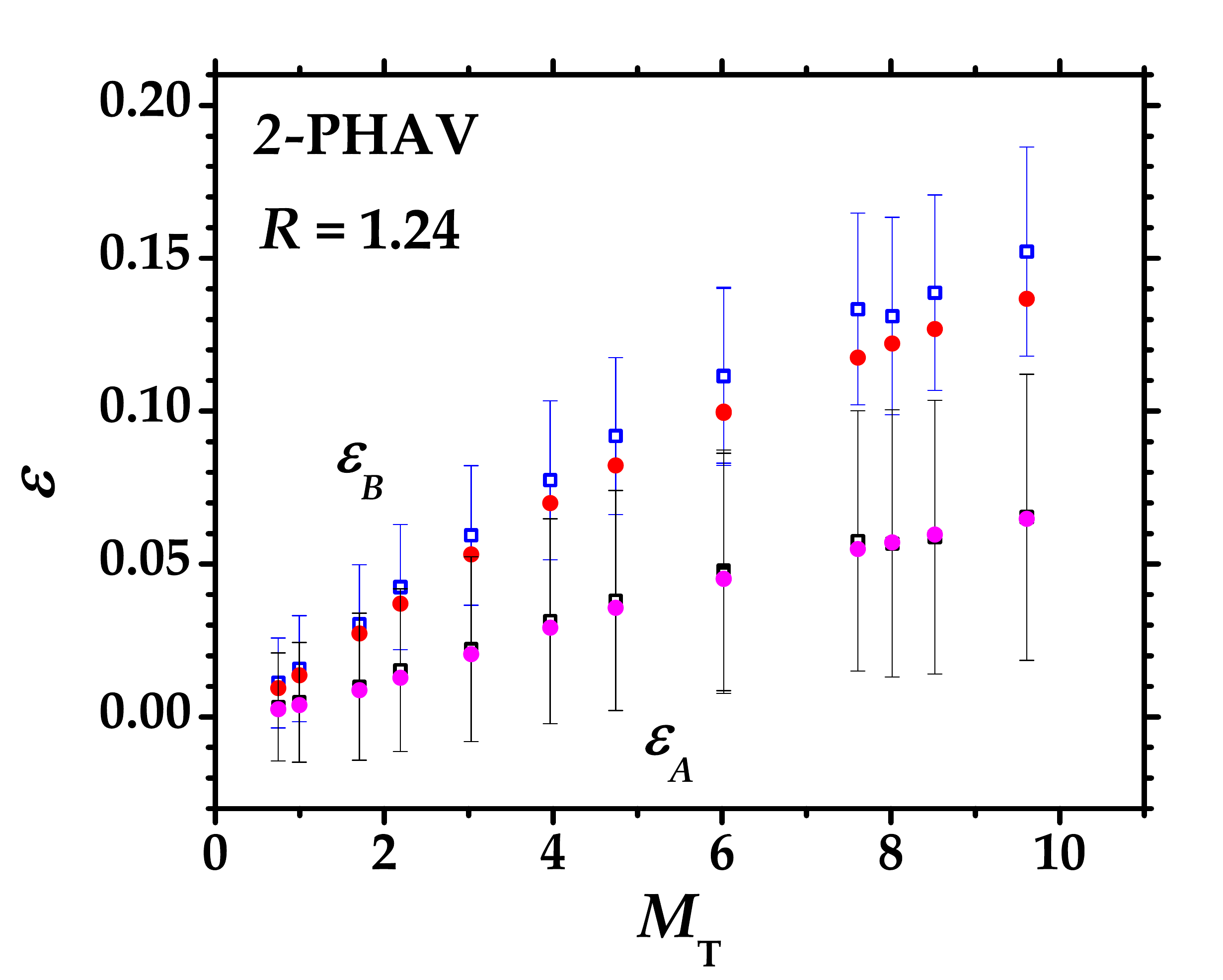} \centering\includegraphics[width=0.49\textwidth]{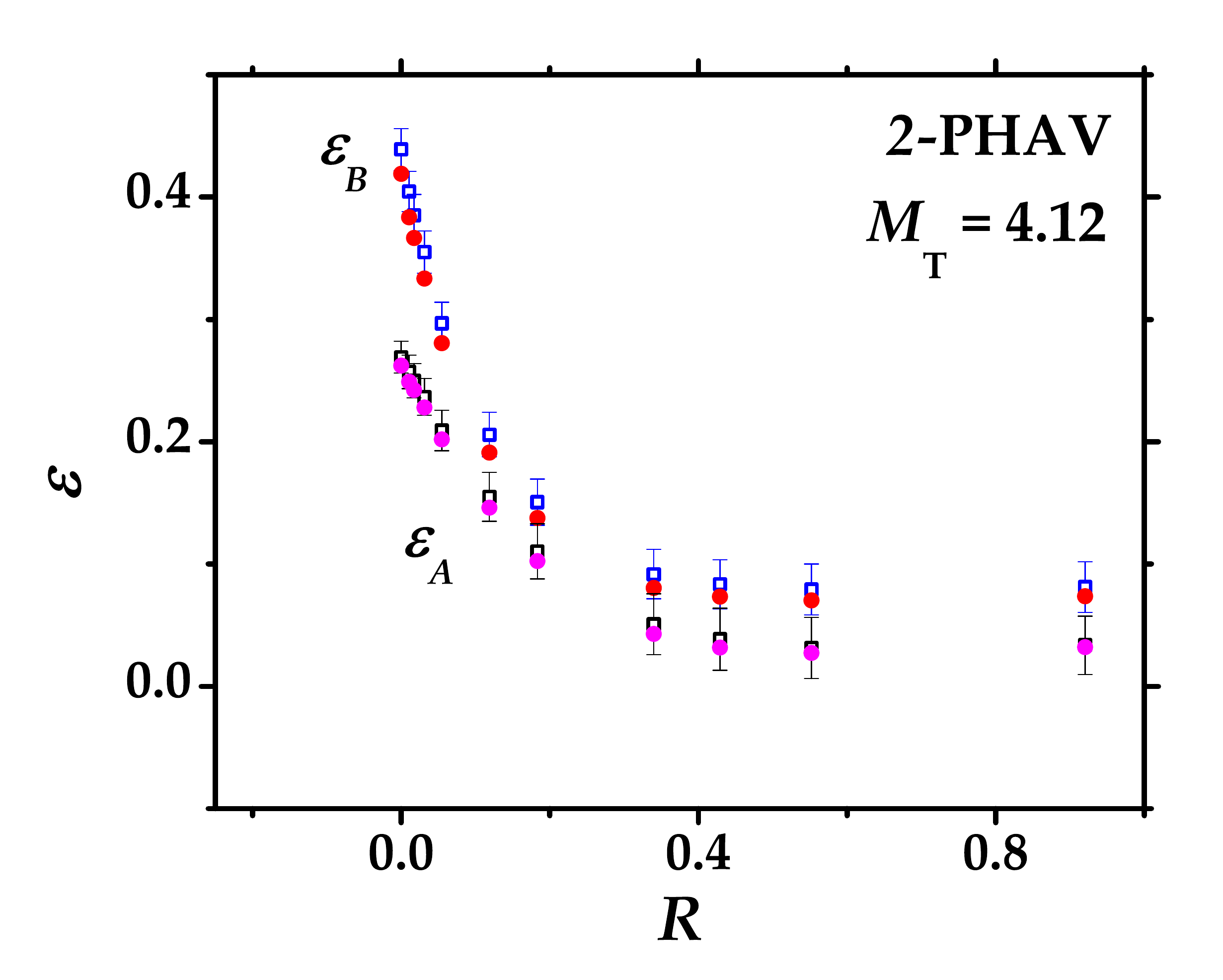} \centering\includegraphics[width=0.49\textwidth]{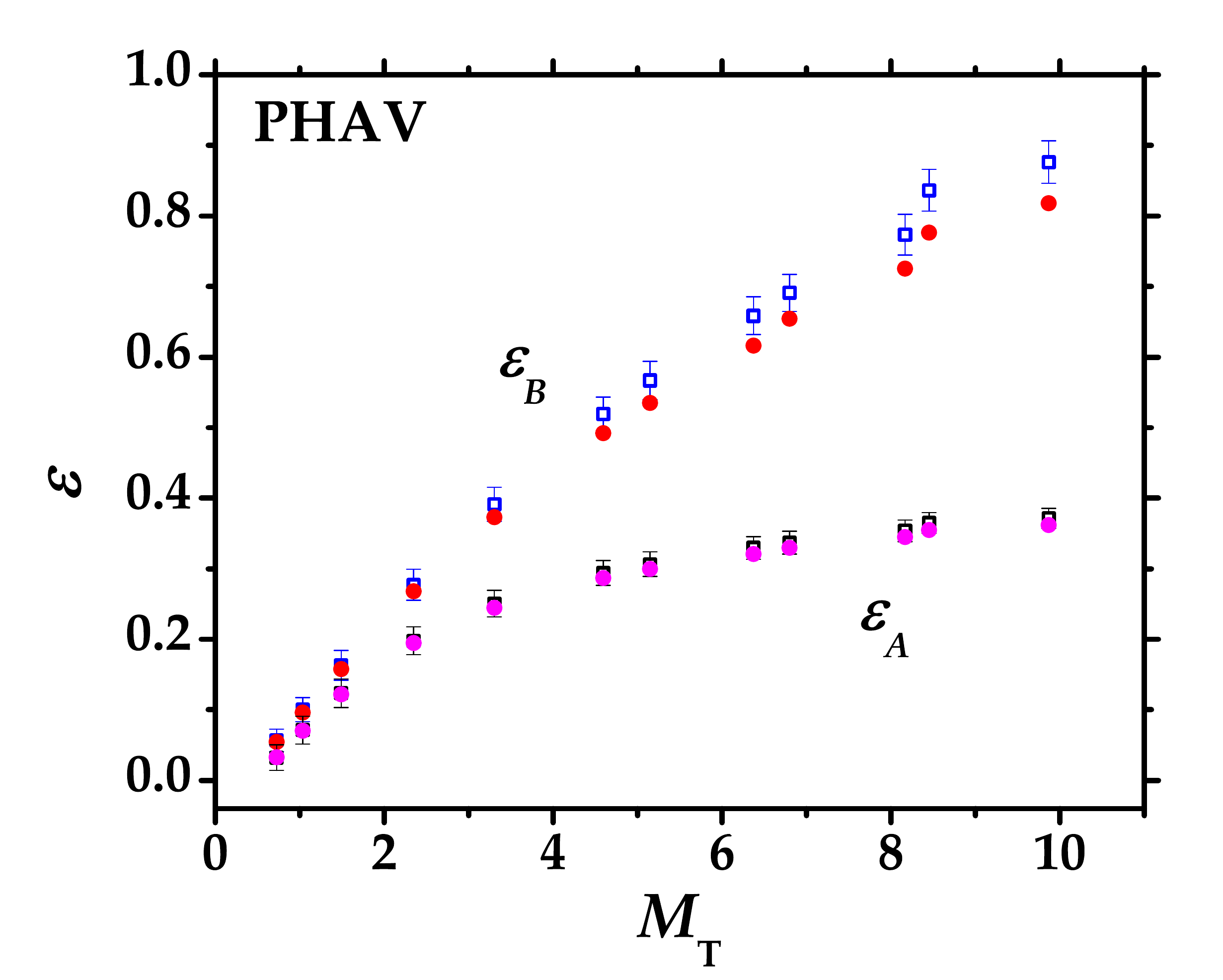}
\caption{Upper left panel: nonG measures $\epsilon_{\rm A}$ and $\epsilon_{\rm B}$ as functions of the mean number of detected photons of almost balanced 2-PHAVs ($R=1.24$).  Upper right panel: $\epsilon_{\rm A}$ and $\epsilon_{\rm B}$ as functions of the balancing between the two components of the 2-PHAV, at fixed mean number of detected photons ($M_T = 4.12$) of the overall state. Lower panel: $\epsilon_{\rm A}$ and $\epsilon_{\rm B}$ as functions of the mean number of detected photons of single PHAVs. Empty symbols: experimental data; Full circles: theoretical expectations.} \label{nonGunbal}
\end{figure}
As 2-PHAV is a state described by 2 parameters, namely the mean value, $M_{\rm T}$, and the balancing $R$ between its two components, we decided to better investigate its nonG nature as a function of one of these variables by keeping fixed the other one. In the upper left panel of Fig.~\ref{nonGunbal} we show $\epsilon_{\rm A}$ and $\epsilon_{\rm B}$ as functions of the mean total energy of the 2-PHAV for a fixed choice of the balancing, namely 1.24: we can notice that the values of both the measures increase at increasing the mean number of detected photons. Moreover, in the upper right panel of the same figure we plot the lower bounds of the nonG measures as functions of the ratio between the two components at fixed mean number of detected photons, that is $M_{\rm T}=4.12$. As one may expect,  it monotonically decreases at increasing the balancing. In fact, the most unbalanced condition reduces to the case in which there is only a single PHAV, whereas the most balanced one corresponds to have a balanced 2-PHAV.
\par
For the sake of completeness, in the lower panel of the same figure we show the results we obtained for the single PHAV as a function of the mean value. Also in this case the experimental data, which are superimposed to the theoretical expectations, testify the accordance between the two measures since they increase their value at increasing the energy of the state.
\par
\begin{figure}[htbp]
\centering\includegraphics[width=0.49\textwidth]{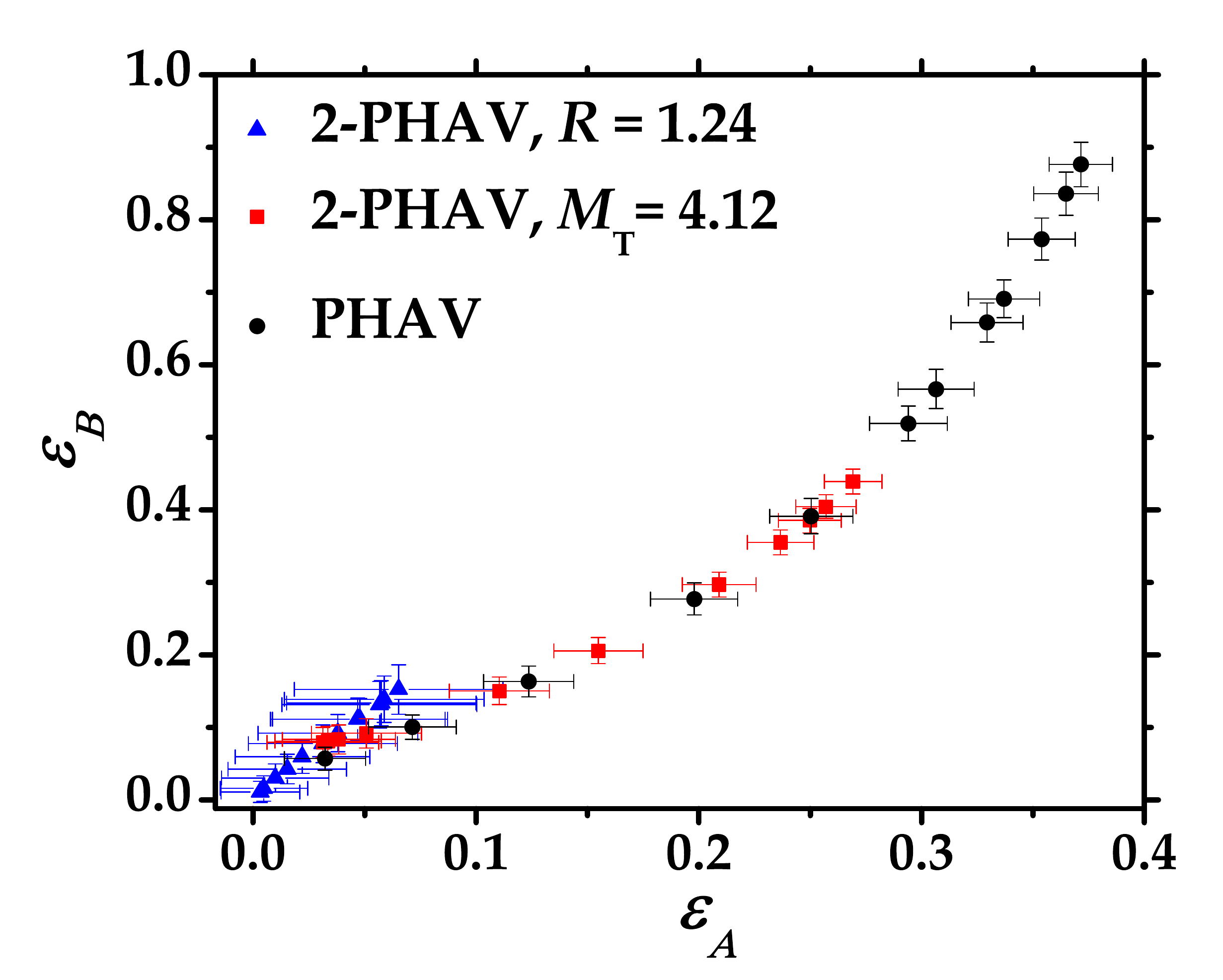} \centering\includegraphics[width=0.49\textwidth]{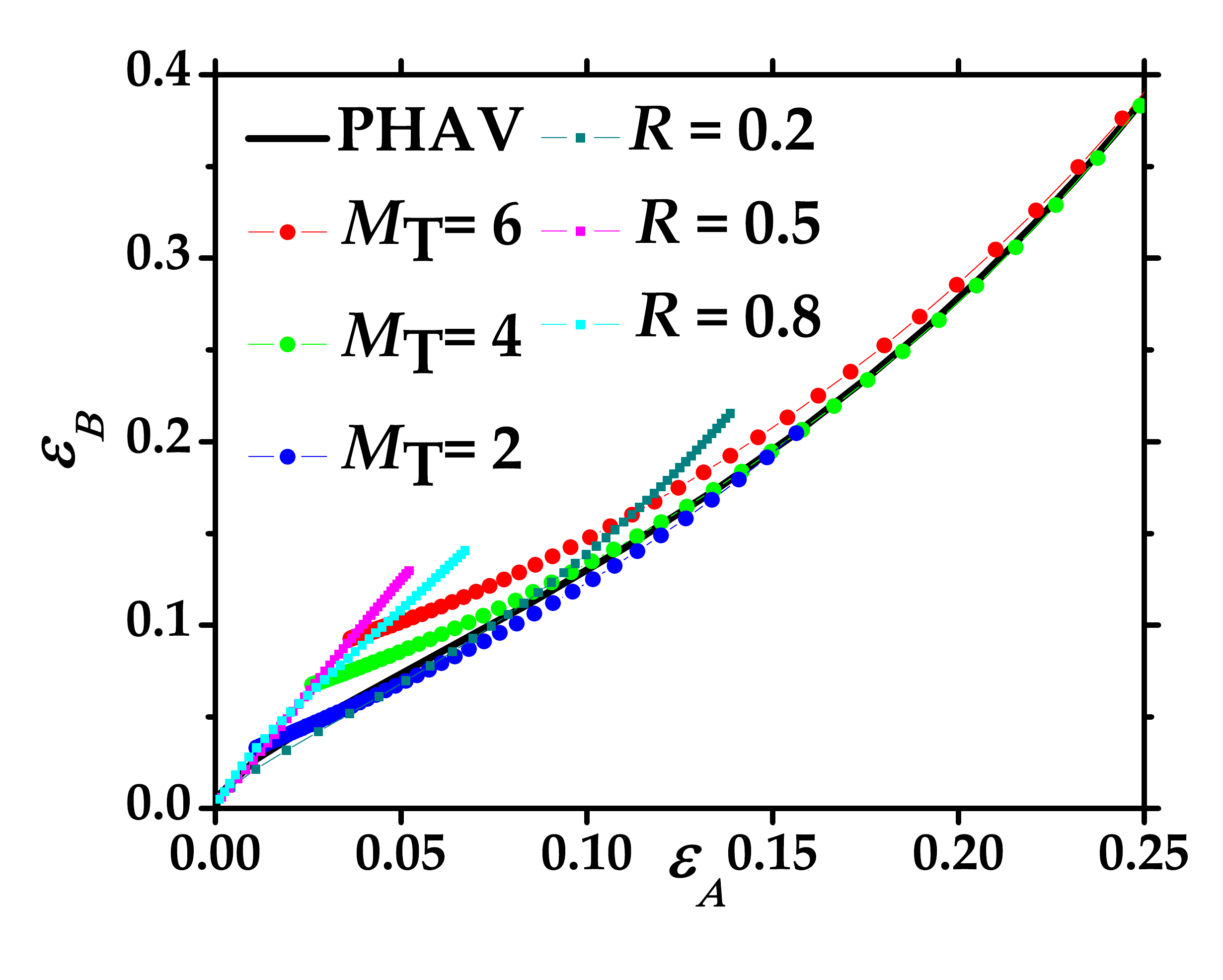}
\caption{Left panel: $\epsilon_{\rm B}$ as a function of $\epsilon_{\rm A}$ for the three experimental cases presented in Fig.~\ref{nonGunbal}. Right panel: simulated behavior of $\epsilon_{\rm B}$ as a function of $\epsilon_{\rm A}$ for different choices of the parameters describing PHAVs and 2-PHAVs (see the text for details).} \label{comparison}
\end{figure}

As in all the cases presented in Fig.~\ref{nonGunbal} the behavior of the two measures is very similar except for the absolute values, we decided to test their monotonicity by following the suggestion of Ref.~[17].
In the left panel of Fig.~\ref{comparison}, we plot the experimental values of measure B as a function of those of measure A for the three cases presented in Fig.~\ref{nonGunbal}. It is evident that the two measures are monotone to each other, even if the absolute values are different. In particular, measure B is endowed with higher values, thus resulting more sensitive to small differences in the choice of the parameters. This property, together with the fact that $\epsilon_{\rm B}$ is characterized by smaller error bars with respect to the other one, can be considered a good criterion to choose one measure instead of the other to quantify the nonG amount and thus discriminate the states under investigation for possible applications.  Nevertheless, it is interesting to notice that the behavior of one amount with respect to the other is not described by a unique curve. In fact, in the case of the 2-PHAV at fixed ratio and variable total energy (blue dots) the curve is slightly different from those corresponding to the other two cases, namely the 2-PHAV at fixed total energy and variable ratio (red dots) or the single PHAV with variable energy (black dots). In order to prove that this difference does not depend on the reliability of the experimental data, in the right panel of Fig.~\ref{comparison} we present the results we obtained by a simulation.    
We plot the theoretical behavior of a single PHAV at different mean numbers of detected photons as black line, whereas we used colored squares + line to indicate the 2-PHAV at fixed ratio ($R=0.2, 0.5, 0.8$) and variable total energy, and colored dots + line to indicate the 2-PHAV at fixed total energy ($M_{\rm T}=2, 4, 6$) and variable ratio. It is evident that there is not a unique curve, as already testified by the experimental data. Nevertheless, we want to notice that there are some limits in which the curves are superimposed (this happens either when the 2-PHAV is almost unbalanced or when it is very low populated as in both the cases it reduces to the case of a single PHAV) or intersect each other (such as in the case in which the 2-PHAV is characterized by a precise choice of total energy and ratio).  
\section{Concluding remarks}\label{s:remarks}
In conclusion, we have presented an experimental investigation of the non-Gaussian nature of the class of PHAVs by reconstructing their Wigner function and using two different measures, both based on quantities experimentally accessed by direct detection, to quantify the nonG amount. We proved the consistency of the different approaches and tested the monotonicity of the two measures.
Nevertheless, the comparison performed on diagonal states belonging to the class of PHAVs for different parameters settings showed that there is not a unique curve describing the behavior of one measure with respect to the other. 
In addition, we discussed the choice of the best measure between the two proposed. According to data, $\epsilon_{\rm B}$ seems to be better because it has higher absolute values and a reduced sensitivity to experimental errors.

\section*{Acknowledgments}
This work has been supported by MIUR (FIRB ``LiCHIS'' - RBFR10YQ3H).

\end{document}